\documentclass[11pt]{article}
\usepackage{amssymb}
\usepackage{amsmath}
\usepackage{graphicx}
\newcommand{\beq}{\begin{equation}}
\newcommand{\eeq}{\end{equation}}
\newcommand{\bea}{\begin{eqnarray}}
\newcommand{\eea}{\end{eqnarray}}
\newcommand{\ba}{\begin{array}}
\newcommand{\ea}{\end{array}}
\newtheorem{theorem}{Theorem}[section]
\newtheorem{prop}[theorem]{Proposition}

\newtheorem{remark}[theorem]{Remark}

\newtheorem{obs}[theorem]{Observation}

\begin{document}

\title{Cohomological, Poisson structures and integrable hierarchies in
tautological subbundles for Birkhoff strata of Sato Grassmannian}
\author{B.G. Konopelchenko \\
Dipartimento di Matematica e Fisica " Ennio de Giorgi", Universit\`{a} del
Salento \\
INFN, Sezione di Lecce, 73100 Lecce, Italy \\
{konopel@le.infn.it} \\
\mbox{} \\
G. Ortenzi \\
Dipartimento di Matematica Pura ed Applicazioni,\\
Universit\`{a} di Milano Bicocca, 20125 Milano, Italy\\
{giovanni.ortenzi@unimib.it} }
\maketitle


\abstract{ \noindent Cohomological and Poisson structures associated with the special tautological
subbundles $TB_{W_{1,2,\dots,n}}$ for the Birkhoff strata of Sato Grassmannian are considered. It is shown that the
tangent bundles of $TB_{W_{1,2,\dots,n}}$ are isomorphic
to the linear spaces of $2-$coboundaries with vanishing Harrison's cohomology modules. Special class of 2-coboundaries
is provided by the systems of integrable quasilinear PDEs. For the
big cell it is the dKP hierarchy. It is demonstrated also that the
families of ideals for algebraic varieties in $TB_{W_{1,2,\dots,n}}$
can be viewed as the Poisson ideals. This observation establishes a
connection between families of algebraic curves in
$TB_{W_{\widehat{S}}}$ and coisotropic deformations of such curves of
zero and nonzero genus described by hierarchies of hydrodynamical
type systems like dKP hierarchy. Interrelation between
cohomological and Poisson structures is noted.}

\section{Introduction}

In this paper we continue the study of algebraic, geometric and other
structures arising in the tautological subbundles for the Birkhoff strata of
Sato Grassmannian. In the paper \cite{KO1} it was shown that each Birkhoff
stratum $\Sigma _{S}$ of the Sato Grassmannian Gr contains a subset $W_{S}$
of points such that for each of these points the corresponding
infinite-dimensional linear space ( fiber of the tautological subbundle $%
TB_{W_{S}}$ associated with $W_{S}$) is closed with respect to pointwise
multiplication. Algebraically all \ $TB_{W_{S}}$ are infinite families of
infinite dimensional associative commutative algebras. Geometrically each
fiber of $TB_{W_{S}}$ is an algebraic variety and the whole $TB_{W_{S}}$ is
an infinite family of algebraic varieties with each finite-dimensional subvariety being a
family of algebraic curves defined by the equations \cite{KO1}

\begin{eqnarray}
p_{j}p_{k}-\sum_{l}C_{jk}^{l}p_{l} &=&0,  \label{alg} \\
\sum_{l}\left( C_{jk}^{l}C_{lm}^{p}-C_{mk}^{l}C_{lj}^{p}\right) &=&0,\qquad
j,k,m,p\in S  \label{ass}
\end{eqnarray}%
where $C_{jk}^{l}$ are parameterized by certain quantities $H_{m}^{n}$.

For the big cell $\Sigma _{\varnothing }$ of the Sato Grassmannian one has $%
C_{jk}^{l}=\delta _{j+k}^{l}+H_{j-l}^{k}+H_{k-l}^{j},j,k=0,1,2,3,\dots $\
and the $TB_{W_{\varnothing }}$ is the collection of families of normal
rational curves (Veronese curves) of the all degrees $2,3,4,\dots $. For the
stratum $\Sigma _{1}$, each fiber of $TB_{W_{1}}$ is the coordinate ring of
the elliptic curve and $\ $ $TB_{W_{1}}$ is the infinite family of such
rings. \ For the set $W_{1,2}$ the $TB_{W_{1,2}}$ is equivalent to the
families of coordinate rings of a special space curve with pretty
interesting properties. This family of curves in $\ TB_{W_{1,2}}$ contains
plane trigonal curve of genus two$.$For the higher strata $\Sigma
_{1,2,\dots ,n}$ $(n=3,4,5,\dots )$ $TB_{W_{1,2,...,n}}$ contains plane $%
(n+1,n+2)$ curve of genus $n$.

\bigskip In the present paper local and Poisson structures associated with
the subbundles $TB_{W_{S}}$ and the corresponding algebraic curves are
discussed. It is shown that the tangent bundles of $TB_{W_{1,2,...,n}}$
carry the hierarchies of integrable equations. In particular, the tangent
subbundle \ $TB_{W_{\varnothing }}$ for the big cell contains the hierarchy
of dispersionless Kadomtsev-Petviashvili ( dKP) equations. It is
demonstrated that the tangent bundles of $TB_{W_{1,2,...,n}}$ and $%
TB_{W_{1,2,...,n}}$ modules $E_{1,2,\dots ,n}$ are isomorphic to the linear
spaces of $2-$coboundaries and Harrison's cohomology modules $H^{2}(W,E)$
and $H^{3}(W,E)$ vanish. Special classes of $2-$cocycles and $2-$%
coboundaries are described by the systems of integrable quasilinear PDEs.
For example, a class of $2-$coboundaries associated with the subbundle $%
TB_{W_{\varnothing }}$ in the big cell is provided by the dKP hierarchy.

We give also an interpretation of the families of ideals $I(\Gamma _{\infty
})$ for families of algebraic curves in $TB_{W_{1,2,...,n}}$ as the Poisson
ideals. It is shown that the family of ideals for the family of normal
rational curves in the big cell is the Poisson ideal with respect to a
Poisson structure obeying certain constraints. Two sets of canonical
variables in such Poisson ideals are used. It is demonstrated that in the
Darboux coordinates the above constraints are nothing else than the dKP
hierarchy. Similar results remain valid for other strata too. Finally an
interrelation between cohomological and Poisson structures of $%
TB_{W_{1,2,...,n}}$ is observed.

The results presented here are the extension of those for Gr$^{(2)}$ (\cite{KO3,KO4}) 
to the general Sato Grassmannian.

The paper is organized as follows. In section 2 we recall the structure of
Sato Grassmannian. Tangent subbundles $TB_{W_{1,2,...,n}}$
and systems of integrable quasilinear PDEs are considered in section 3.
In section 4 we study connection between Harrison cohomology and dKP hierarchy. 
Associated Poisson ideals and coisotropic deformations are discussed in
section 4.

\section{Sato Grassmannian, Birkhoff strata and algebraic varieties}

Here we briefly recall, without entering in any technicalities, some basic
facts about  Sato Grassmannian, its Birkhoff stratification ( see e.g., in 
\cite{PS,SW})and results obtained in \cite{KO1}. The Sato Grassmannian Gr can
be viewed as the set of all subspaces of the infinite-dimensional set of all
formal Laurent series with complex-valued coefficients admitting an
algebraic basis ($w_{0}(z),$ $w_{1}(z),...)$ with the basis elements

\begin{equation}
w_{n}=\sum_{k=-\infty }^{n}H_{k}^{n}z^{k}  \label{bas}
\end{equation}%
of finite order n. The Grassmanian Gr has a stratified structure $Gr=\cup
_{s}\Sigma _{s}$ where the strata \ $\Sigma _{s}$ are subsets in Gr which
are span of \ basis elements (\ref{bas}) with particular set of integers $n$ (see \cite{PS,SW}). For the big cell the basis is composed by formal Laurent
series 
\begin{equation}
p_{0}=1,\qquad p_{i}(z)=z^{i}+\sum_{k=1}^{\infty }\frac{H_{k}^{i}}{z^{k}}%
,\qquad i=1,2,\dots .  \label{bigcellbasis}
\end{equation}%
This set is identified by the infinite set of symbols $H_{j}^{i}$, $i,j\geq 1
$. The big cell
is a dense set in the Sato Grassmannian. The points outside the big cell form
Birkhoff strata. It can be shown that the points in the Birkhoff strata can
be constructed changing the basis of the space of Laurent series and, in a
suitable way (\cite{PS,SW}), the queue of the Laurent series. In the simplest
case of first Birkhoff stratum $\Sigma _{1}$ the basis of Laurent series is
obtained removing the element $p_{1}$ and hence it is of the form 
\begin{equation}
p_{0}=1,\qquad p_{i}(z)=z^{i}+H_{-1}^{i}z+\sum_{k=1}^{\infty }\frac{H_{k}^{i}%
}{z^{k}},\qquad i=2,3\dots .  \label{1sbasis}
\end{equation}%
plus element of the degree $-1$. As in $\Sigma _{\varnothing }$ case, a point
of $\Sigma _{1}$ is a span of the Laurent series $\{p_{i}\}_{-1,0,2,3,\dots }
$.

In this paper, following \cite{KO1}, will be interested to a particular
subsets $W_{S}$ of the Birkhoff strata of the Sato Grassmannian. In general,
$W_{S}$ are composed by the Laurent series  closed with respect to
point-wise multiplication. In the case of the big cell $W_{\varnothing }$ is
given by the span of the elements \ref{bigcellbasis} closed with respect to
multiplication, i.e. such that 
\begin{equation}
p_{j}(z)p_{k}(z)=\sum_{l=0}C_{jk}^{l}p_{l}(z),\qquad j,k=0,1,2,\dots 
\label{bigcellalg}
\end{equation}%
for some suitable $C_{jk}^{l}$. This property is verified iff  the
coefficients $H_{j}^{i}$ of the formal Laurent series $p_{i}$ obey the
constraints \cite{KO1} 
\begin{equation}
H_{m}^{j+k}-H_{m+k}^{j}-H_{j+m}^{k}+\sum_{l=1}^{j-1}H_{j-l}^{k}H_{m}^{l}+%
\sum_{l=1}^{k-1}H_{k-l}^{j}H_{m}^{l}-\sum_{l=1}^{m-1}H_{m-l}^{k}H_{l}^{j}=0,%
\qquad j,k,m=1,2,3,\dots .  \label{bcsc}
\end{equation}%
and  
\begin{equation}
C_{jk}^{l}=\delta _{j+k}^{l}+H_{j-l}^{k}+H_{k-l}^{j},\qquad
j,k=0,1,2,3,\dots \ .  \label{bcstrc}
\end{equation}%
Under these constraints the Laurent series $p_{i}$ can be presented as  
\begin{equation}
\begin{split}
p_{0}=& 1, \\
p_{2}=& {p_{1}}^{2}-2H_{1}^{1}, \\
p_{3}=& {p_{1}}^{3}-3H_{1}^{1}p_{1}-3H_{2}^{1}, \\
p_{4}=& {p_{1}}^{4}-4H_{1}^{1}{p_{1}}^{2}-4H_{2}^{1}p_{1}-4H_{3}^{1}+2{%
H_{1}^{1}}^{2}, \\
p_{5}=& {p_{1}}^{5}-5H_{1}^{1}{p_{1}}^{3}-5H_{2}^{1}{p_{1}}^{2}-\left( 5\,H_{%
{3}}^{1}-5\,{H_{{1}}^{1}}^{2}\right) p_{1}-5\,H_{{4}}^{1}+5\,H_{{1}}^{1}H_{{2%
}}^{1}, \\
\dots & \ . \\
&
\end{split}
\label{bigcellcurr}
\end{equation}%
One has similar results for all the Birkhoff strata. For the first stratum 
$\Sigma _{1}$ the subset of Laurent series closed with respect to
point-wise product $p_{i}p_{j}=\sum_{k}C_{ij}^{k}p_{k}$ can be constructed
iteratively. In this case  the coefficients $H_{j}^{i}$ should satisfy the
constraints 
\begin{equation}
\begin{split}
H_{j+l}^{i}&
+H_{i+l}^{j}+H_{-1}^{j}H_{l+1}^{i}+H_{-1}^{i}H_{l+1}^{j}+%
\sum_{n=1}^{l-1}H_{n}^{j}H_{l-n}^{i}= \\
&
H_{l}^{i+j}+H_{-1}^{j}H_{l}^{i+1}+H_{-1}^{i}H_{l}^{j+1}+%
\sum_{n=2}^{i-1}H_{i-n}^{j}H_{l}^{n}+%
\sum_{n=2}^{j-1}H_{j-n}^{i}H_{l}^{n}+H_{-1}^{i}H_{-1}^{j}H_{l}^{2}+ \\
& (H_{j}^{i}+H_{i}^{j}+H_{-1}^{i}H_{1}^{j}+H_{1}^{i}H_{-1}^{j})\delta
_{0}^{l},\qquad j,k=2,3,4,\dots ,\ l=-1,1,2,3,\dots .
\end{split}
\label{1stratsercoeff}
\end{equation}%
and 
\begin{equation}
C_{ij}^{l}=\delta _{i+j}^{l}+H_{-1}^{j}\delta _{i+1}^{l}+H_{-1}^{i}\delta
_{j+1}^{l}+H_{i-l}^{j}+H_{j-l}^{i}+H_{-1}^{i}H_{-1}^{j}\delta
_{2}^{l}+\left(
H_{j}^{i}+H_{i}^{j}+H_{-1}^{i}H_{1}^{j}+H_{1}^{i}H_{-1}^{j}\right) \delta
_{0}^{l}.  \label{1sstructcoeff}
\end{equation}%
The first $p_{i}$'s are 
\begin{equation}
\begin{split}
p_{4}=& {p_{{2}}}^{2}-2\,{H^{2}}_{{-1}}p_{{3}}-{{H^{2}}_{{-1}}}^{2}p_{{2}%
}-2\,{H^{2}}_{{2}}-2\,{H^{2}}_{{-1}}{H^{2}}_{{1}} \\
p_{5}=& p_{{2}}p_{{3}}-{H^{2}}_{{-1}}{p_{{2}}}^{2}-\left( {H^{3}}_{{-1}}-2\,{%
{H^{2}}_{{-1}}}^{2}\right) p_{{3}}-\left( {H^{2}}_{{1}}+{H^{2}}_{{-1}}{H^{3}}%
_{{-1}}-{{H^{2}}_{{-1}}}^{3}\right) p_{{2}} \\
& -\frac{3}{2}\,{H^{2}}_{{1}}{H^{3}}_{{-1}}-\frac{5}{2}\,{H^{2}}_{{3}}-1/2\,{%
H^{2}}_{{-1}}{H^{3}}_{{1}}+2\,{H^{2}}_{{-1}}{H^{2}}_{{2}}+2\,{{H^{2}}_{{-1}}}%
^{2}{H^{2}}_{{1}} \\
\dots & .
\end{split}
\label{1scur}
\end{equation}%
The analogue of the subset $W_{\varnothing }$ for $\Sigma _{1}$ is given by
the points of $\Sigma _{1}$ whose $p_{i}$'s satisfy (\ref{1scur},\ref%
{1sstructcoeff}). The relations (\ref{bigcellalg}) naturally defines a set
of quadric algebraic varieties in an infinite dimensional space of
coordinates $p_{i}$. Every variety is of genus zero. This property changes
in the Birkhoff strata. In the space of coordinates $p_{0},p_{2},p_{3},\dots 
$ there are an infinite number of quadric algebraic varieties defined by the
relations $p_{i}p_{j}=\sum_{k}C_{ij}^{k}p_{k}$. For $W_{\Sigma_1}$ there is an elliptic (genus $%
1$) curve 
\begin{equation}
\begin{split}
\mathcal{F}_{23}^{1}=& {p_{3}}^{2}-{p_{2}}^{3}+3\,H_{-1}^{2}p_{3}\,p_{2}-2%
\,H_{{-1}}^{3}{p_{2}}^{2}+\left( {H_{{-1}}^{2}}^{3}+3\,H_{{1}}^{2}+H_{{-1}%
}^{2}H_{{-1}}^{3}\right) p_{3} \\
& -\left( {H_{{-1}}^{3}}^{2}+2\,H_{{1}}^{3}-3\,H_{{-1}}^{2}H_{{1}}^{2}-3\,H_{%
{2}}^{2}+H_{{-1}}^{3}{H_{{-1}}^{2}}^{2}\right) {p_{2}} \\
& -2\,H_{{3}}^{3}-2\,H_{{-1}}^{3}H_{{1}}^{3}+3\,H_{{4}}^{2}+3\,H_{{2}}^{2}{%
H_{{-1}}^{2}}^{2}-\frac{3}{2}\,H_{{-1}}^{2}H_{{\ 3}}^{2}+3\,{H_{{1}}^{2}}%
^{2}-\frac{3}{2}\,{H_{{-1}}^{2}}^{2}H_{{1}}^{3} \\
& -\frac{1}{2}\,H_{{-1}}^{2}H_{{1}}^{2}H_{{-1}}^{3}+4\,H_{{-1}}^{3}H_{{2}%
}^{2}=0.
\end{split}
\label{1stratellcurv-red}
\end{equation}%
among them \cite{KO1}.

\section{Tangent subbundles $TB_{S}$ and systems of integrable quasilinear
PDEs}

In this paper we will consider only the subsets $W_{1,2,...,n}$ of the
Birkhoff strata and the corresponding subbundles $TB_{W_{1,2,...,n}}$. A
standard method to analyze local properties of the varieties defined by
equations (\ref{alg},\ref{ass}) is to deal with their tangent bundle $T_{W}$
(\cite{HP}-\cite{Har}). Let us denote by $\pi _{i}$ and $\Delta _{ik}^{l}$
the corresponding elements of $T_{W}$ in a point. They are defined, as
usual, by the system of linear equations 
\begin{eqnarray}
&&\pi _{j}p_{k}+p_{j}\pi _{k}-\sum_{l}\Delta
_{jk}^{l}p_{l}-\sum_{l}C_{jk}^{l}\pi _{l}=0,  \label{Talg} \\
&&\sum_{l}\left( \Delta _{jk}^{l}C_{lm}^{p}+C_{jk}^{l}\Delta
_{lm}^{p}-\Delta _{mk}^{l}C_{lj}^{p}-C_{mk}^{l}\Delta _{lj}^{p}\right)
=0,\qquad j,k,m,p\in S_{1,2,\dots ,n}.  \label{Tass}
\end{eqnarray}%
In more general setting these equations define also a $TB_{W_{1,2,...,n}}$%
-module $E$. 

For the Birkhoff strata the structure constants $C_{jk}^{l}$ have a special
structure being parameterized by $H_{k}^{j}$. Consequently $\Delta _{jk}^{k}$
are also parameterized by $\Delta _{jk}$, i.e. by images of $H_{k}^{j}$ in
the map $W\rightarrow E$, and equations (\ref{ass}) becomes linear equations
(\ref{Tass}) for $\Delta _{jk}$. Being the elements of $E$ (in particular,
the tangent space in a point) $\Delta _{ij}$ admit a natural Ansatz 
\begin{equation}
\Delta _{jk}=\frac{\partial u_{k}}{\partial x_{j}}  \label{D-ans}
\end{equation}%
where $u_{k}$ is a set of independent coordinates for the variety defined by
the associativity condition (\ref{ass}) and $x_{j}$ is a set of new
independent parameters. Under the Ansatz (\ref{D-ans}) equations (\ref{Tass}%
) take the form of quasilinear PDEs for $u_{i}$. 

These systems of quasilinear PDEs are very special. Let us consider the
subbundle $TB_{W_{\varnothing }}$ for the big cell. Since in this case 
\begin{equation}  \label{bigcellstructcoeff}
C_{jk}^{l}=\delta _{j+k}^{l}+H_{j-l}^{k}+H_{k-l}^{j},
\end{equation}
and 
\begin{equation}  \label{bigcellsercoeff}
H^{j+k}_m+H^{j}_{m+k}-H^k_{j+m}+\sum_{l=1}^{j-1}
H^k_{j-l}H^l_m+\sum_{l=1}^{k-1} H^j_{k-l}H^l_m -\sum_{l=1}^{m-1}
H^k_{m-l}H^j_l=0, \qquad j,k,m=1,2,3,\dots,
\end{equation}
one has $\Delta_{kj}^{l}=\Delta _{k,j-l}+\Delta _{j,k-l}$ and the system (%
\ref{Tass}) is 
\begin{equation}
\begin{split}
\Delta _{j+k,m}& +\left( -\Delta _{j,m+k}+\sum_{l=1}^{j-1}H_{j-l}^{k}\Delta
_{lm}+\sum_{l=1}^{k-1}H_{k-l}^{j}\Delta
_{lm}-\sum_{l=1}^{m-1}H_{m-l}^{k}\Delta _{jl}\right) \\
& +\left( -\Delta _{k,m+j}+\sum_{l=1}^{j-1}\Delta
_{j-l}^{k}H_{lm}+\sum_{l=1}^{k-1}\Delta
_{k-l}^{j}H_{lm}-\sum_{l=1}^{m-1}\Delta _{m-l}^{k}H_{jl}\right) =0.
\end{split}
\label{bigcelllin}
\end{equation}%
This system implies 
\begin{equation}
k\Delta _{ik}-i\Delta _{ki}=0.  \label{bigcellsymmetrizz}
\end{equation}%
Let us rename $H_{k}^{1}$ as $u_{k}$.

\begin{prop}
\label{prop-bigcellansatz} Under the Ansatz 
\begin{equation}  \label{bigcellansatz}
\Delta_{ik}=\frac{\partial u_k}{\partial x_i}, \qquad i,k=1,2,3,\dots\ .
\end{equation}
the system (\ref{bigcelllin}) coincides with the dKP hierarchy.
\end{prop}

\textbf{Proof } For $j=1,k=2,m=1$ the system (\ref{bigcelllin}) is 
\begin{equation}
\Delta _{31}-\Delta _{13}-\Delta _{22}+2H_{1}^{1}\Delta _{11}=0
\end{equation}%
while the relations (\ref{bigcellsymmetrizz}) at $i=1,k=2$ and $i=1,k=3$ are 
\begin{equation}
2\Delta _{12}-\Delta _{21}=0,\qquad 3\Delta _{13}-\Delta _{31}=0.
\end{equation}%
The ansatz (\ref{bigcellansatz}) gives 
\begin{equation}
\begin{split}
& \partial _{x_{3}}u_{1}-\frac{3}{2}\partial
_{x_{2}}u_{2}+3u_{1}\partial_{x_{1}}u_{1}=0, \\
& 2\partial _{x_{1}}u_{2}-\partial _{x_{2}}u_{1}=0.
\end{split}
\label{dKP-KZ}
\end{equation}%
It is the celebrated dKP (Khokhlov-Zaboloskaya) equations (see e.g. \cite%
{Zak1}-\cite{Kri5},\cite{KM}). Similarly 
for $j=1$, $k=1$ and $m=3$ the system (\ref{bigcelllin}) is  
\begin{equation}
\Delta_{23}-2\Delta_{14} -2 H_2^1 \Delta_{11} -2 H_1^1 \Delta_{12} =0
\end{equation}
and the relations (\ref{bigcellsymmetrizz}) at $i=1,k=4$, and $i=1,k=3$ are 
\begin{equation}
\Delta_{41} = 4 \Delta_{14}, \qquad 3 \Delta_{13} = \Delta_{31}.
\end{equation}
These equations, using also (\ref{dKP-KZ}), give 
\begin{equation}
\begin{split}
& \partial_{x_4} u_1=2\partial_{x_2} u_3 -2 \partial_{x_1} (u_1 u_2) \\
& \partial_{x_1} u_3 = \frac{1}{2}\partial
_{x_{2}}u_{2}-u_{1}\partial_{x_{1}}u_{1} \\
& 2\partial _{x_{1}}u_{2}-\partial _{x_{2}}u_{1}=0
\end{split}%
\end{equation}
which the second equation in the dKP hierarchy. The higher equations (\ref%
{bigcelllin}), (\ref{bigcellsymmetrizz}) give rise to the higher dKP
equations under the ansatz (\ref{bigcellansatz}). $\square $

The Ansatz (\ref{D-ans}) is necessary and sufficient condition for the
closeness of the differential one-forms \ $\Omega _{k}=\sum_{j=0}^{\infty
}\Delta _{jk}dx_{j},k=1,2,...$. Hence, the above observation can be
formulated also in the following form.

\begin{prop}
Special subbundle $TB_{W_{\varnothing }}$ for which differential one-forms $%
\Omega _{k}=\sum_{j=0}^{\infty }\Delta _{jk}dx_{j},k=1,2,...$ are closed is
governed by the dKP hierarchy.
\end{prop}

\section{Harrison cohomology and dKP Harrison cohomology}

Equations (\ref{Talg}) and (\ref{Tass}) imply certain cohomological
properties of the subbundles $TB_{W_{1,2,...,n}}$. Indeed, if one introduces
the bilinear map $\psi (\alpha ,\beta )$ with $\alpha ,\beta \in $ $%
TB_{W_{1,2,...,n}}$defined by (see e.g. \cite{Sha}) 
\begin{equation}
\psi (p_{i},p_{k})=\sum_{l}\Delta _{jk}^{l}p_{l}.
\end{equation}%
Then the equations (\ref{Tass}) take the form

\begin{equation*}
p_{j}\psi (p_{k},p_{l})- \psi(p_j p_k, p_l ) + \psi(p_j, p_k p_l ) - p_l \psi (p_j, p_k ) =0,
\end{equation*}%
or equivalently

\begin{equation}
\alpha \psi (\beta ,\gamma )-\psi (\alpha \beta ,\gamma )+\psi (\alpha
,\beta \gamma )-\gamma \psi (\alpha ,\beta )=0  \label{2cocy}
\end{equation}%
where $\alpha ,\beta ,\gamma \in $.$TB_{W_{1,2,...,n}}$ Bilinear maps of
such type are called Hochschild $2-$cocycles \cite{Hoc}. So, the tangent
bundle to the variety of the structure constants $C_{jk}^{l}$ is isomorphic
to the linear space of the $2-$cocycles on $\ TB_{W_{1,2,...,n}}$(see e.g. 
\cite{Sha}). For the \ commutative algebras this classical results
represents a part of the cohomology theory of commutative associative
algebras proposed by Harrison in \cite{Hron}. It is the most appropriate
tool to analyze the local properties of the varieties $W_{s}$. 

Equations (\ref{Talg}) gives us an additional information about the $2-$%
cocycle $\psi (\alpha ,\beta )$. Introducing a linear map $g(\alpha )$
defined by $g(p_{i})=\pi _{i}$, one rewrites equation (\ref{Talg}) as

\begin{equation*}
p_{j}g(p_{k})+p_{k}g(p_{j})-\psi (p_{j},p_{k})-g(p_{j}p_{k})=0
\end{equation*}%
Thus, 
\begin{equation}
\psi (\alpha ,\beta )=\alpha g(\beta )+\beta g(\alpha )-g(\alpha \beta )
\label{1cob}
\end{equation}%
with $\alpha ,\beta \in TB_{W_{1,2,...,n}}$. So 
\begin{equation}
\psi (\alpha ,\beta )=\delta g(\alpha ,\beta )
\end{equation}%
where $\delta $ is the Hochschild coboundary operation. Hence, $\psi (\alpha
,\beta )$ is a $2-$coboundary and one has

\begin{prop}
\label{prop-coc-cob} The tangent bundle of the subbundle $TB_{W_{1,2,...,n}}$
is isomorphic to the linear space of $2-$coboundaries and Harrison's
cohomology modules $H^{2}(TB_{W_{1,2,...,n}},E)$ and $%
H^{3}(TB_{W_{1,2,...,n}},E)$ vanish.
\end{prop}

This statement is essentially the reformulation for the subbundles $%
TB_{W_{1,2,...,n}}$of the well-known results concerning the cohomology of
commutative associative algebras (see e.g. \cite{Hron}-\cite{Fro1}). In
particular the existence of the $2-$cocycle and $%
H^{2}(TB_{W_{1,2,...,n}},E)=0$ is sufficient condition for the regularity of
the point at which it is calculated (see e.g. \cite{Hron}-\cite{NR}).

The above results are valid for all algebraic varieties associated with the
Birkhoff strata. The observation made in the previous section shows that
among the generic 2-coboundaries considered above there are special ones
associated with the integrable equations. For the big cell, as the 
immediate consequence of the Proposition \ref{prop-bigcellansatz}, one has

\begin{prop}
Solutions of the dKP hierarchy provide us with the class of $2-$cocycles and 
$2-$ coboundaries defined by 
\begin{equation}
\psi (p_{j},p_{k})=\sum_{l}\left( \frac{\partial u_{j-l}}{\partial x_{k}}+%
\frac{\partial u_{k-l}}{\partial x_{j}}\right) p_{l}  \label{dKP-cob}
\end{equation}%
for the subbundle $TB_{W_{\varnothing }}$ in the big cell.
\end{prop}

We will refer to such $2-$coboundaries as dKP $2-$coboundaries. These dKP $%
2- $coboundaries describe local properties of the family of normal rational
curves.

In terms of the dKP tau-function $F$ defined by (see e.g. \cite{TT,KM}) 
\begin{equation}
u_k=H^1_k=-\frac{1}{k}\frac{\partial F}{\partial x_1\partial x_k}
\end{equation}
the whole dKP hierarchy is represented by the celebrated dispersionless
Hirota-Miwa equations (see e.g. \cite{TT},\cite{KM}) 
\begin{gather}
\begin{aligned}\label{Hirota}
&-\frac{1}{m}F_{i+k,m}+\frac{1}{m+k}F_{i,k+m}+\frac{1}{i+m}F_{k,i+m}
+\sum_{l=1}^{i-1}\frac{1}{m(i-l)}F_{k,i-l}F_{l,m}\\&+\sum_{l=1}^{k-1}%
\frac{1}{m(k-l)}F_{i,k-l}F_{l,m}
-\sum_{l=1}^{m-1}\frac{1}{i(m-l)}F_{k,m-l}F_{i,l}=0 \end{aligned}
\end{gather}
where $F_{i,k}$ stands for the second-order derivative of $F$ with respect
to $x_{i}$ and $x_{k}$. So any solution $F$ of the system (\ref{Hirota})
provides us with the dKP $2-$cocycles (and $2-$coboundaries) given by 
\begin{equation}
\psi(p_j,p_k)=-\sum_{l=1}\left(\frac{1}{j-l}\frac{\partial^2}{\partial x_k
\partial x_{j-l}} +\frac{1}{k-l}\frac{\partial^2}{\partial x_j \partial
x_{k-l}}\right)\frac{\partial F}{\partial x_1} p_l.
\end{equation}
This formula shows that the choice (\ref{D-ans}) corresponds to a simple
realization of the map $W_\varnothing \to E$, namely, $F \to \frac{\partial F%
}{\partial x_1}$ or $H^j_k \to \frac{\partial H^j_k}{\partial x_1}$.

It is evident that all above expressions are well defined only for bounded $%
\frac{\partial u_i}{\partial x_k}$. When $\frac{\partial u_i}{\partial x_k}
\to \infty$ the formulas presented above break down and $H^2(W,E) \neq 0$.

For the dKP hierarchy the points where $\frac{\partial u_i}{\partial x_k}
\to \infty$ are the, so-called, breaking points (or points of gradient
catastrophe). Such points form the singular sector of the space of solutions
of the dKP hierarchy. In this sector the space of variables $x_1,x_2,\dots$
is stratified and such stratification is closely connected with the Birkhoff
stratification. For Burgers-Hopf hierarchy ($2-$reduction of the dKP
hierarchy) and the Grassmannian Gr$^{(2)}$ such situation has been analyzed
in \cite{KK}.

We are confident that similar results hold for other strata too. A complete
analysis of the Harrison cohomology of the tautological subbundles for
Birkhoff strata of the Grassmannian $Gr^{(2)}$ and corresponding integrable
equations has been performed in \cite{KO3}

\section{Families of curves, Poisson ideals and coisotropic deformations}

Families of curves, algebraic varieties and families of their ideals
considered above can be viewed also as embedded in larger spaces with
certain specific properties, for instance, as the coisotropic submanifolds
of Poisson manifolds and Poisson ideals, respectively. Recall that a
submanifold in the Poisson manifold equipped with the Poisson bracket $\{\
,\ \}$ is a coisotropic submanifold if its ideal $\mathcal{I}$ is the
Poisson ideal (see e.g \cite{Wei}), i.e. 
\begin{equation}  \label{IdId=Id}
\{ \mathcal{I} , \mathcal{I} \} \subset \mathcal{I}.
\end{equation}
Relevance of Poisson ideals in the study of (quantum) cohomology of
manifolds was observed in the paper \cite{GK}. Theory of coisotropic
deformations of commutative associative algebras based on the property (\ref%
{IdId=Id}) has been proposed in \cite{KM}. An extension of this theory to
general algebraic varieties was given in \cite{KO}.

Thus let us consider an infinite-dimensional Poisson manifold $P$ with local
coordinates $q_1$, $q_2$, $q_3$, $\dots$, $y_1$, $y_2$, $y_3$, $\dots$
endowed with the Poisson bracket defined by the relations 
\begin{equation}  \label{py=J}
\{q_i,q_k\}=0, \quad \{y_i,y_k\}=0, \quad \{y_i,q_k\}=J_{ki}, \qquad
i,k=1,2,3,\dots
\end{equation}
where $J_{ki}$ are certain functions of $p$ and $y$. This choice of the
Poisson structure is suggested by the roles that the variables $p_i$ and $%
y_k $ play in our construction. Jacobi identities for the Poisson structures
(\ref{py=J}) are given by the system 
\begin{equation}  \label{bigcell-Jacobi}
\begin{split}
&\sum_s J_{ls} \partial_{y_s} J_{kj}-\sum_s J_{ks} \partial_{y_s} J_{lj}=0,
\\
&\sum_s J_{sj} \partial_{q_s} J_{lk}-\sum_s J_{sk} \partial_{q_s} J_{lj}=0.
\end{split}%
\end{equation}
Then, we consider ideals $\mathcal{I}(\Gamma_\infty)$ of the families of
algebraic varieties in $W_{1,2,\dots,n}$ as ideals in $P$ and require that
they are Poisson ideals, i.e. subalgebras
\begin{equation}  \label{IGIG=IG}
\{ \mathcal{I}(\Gamma_\infty) , \mathcal{I}(\Gamma_\infty) \} \subset 
\mathcal{I}(\Gamma_\infty).
\end{equation}
The property (\ref{IGIG=IG}) means, in particular, that the Hamiltonian
vector fields generated by each member of $\mathcal{I}(\Gamma_\infty)$ are
tangent to the coisotropic submanifold with the ideal $\mathcal{I}%
(\Gamma_\infty)$.

The crucial question now is whether a Poisson structure exists such ideals $%
\mathcal{I}(\Gamma_\infty)$ obey (\ref{IGIG=IG}). Let us begin with the big
cell. For the subbundle $TB_{W_\varnothing}$ the answer is given by

\begin{prop}
The family of ideals $I(\Gamma_\infty)$ of the family of normal rational
curves in the big cell represents the Poisson ideal in the Poisson manifold
endowed with the Poisson brackets (\ref{py=J}) with $J_{ik}$ obeying the
constraints 
\begin{equation}  \label{bigcellJcond}
(J_{i\ k-1}-J_{k\ i-1})|_{\Gamma_\infty}=0\qquad i,k-2,3,4,\dots\ .
\end{equation}
\end{prop}

\textbf{Proof } To prove (\ref{IGIG=IG}) it is sufficient to show that for
the elements $h_n$ of the basis of $I(\Gamma_\infty)$ one has $%
\{h_n,h_m\}\subset I(\Gamma_\infty)$. The local coordinates $p^*_n=P_n(-p_1,-%
\frac{1}{2}p_2,-\frac{1}{3}p_3,\dots)$, $n=2,3,4,\dots$, where $%
P_n(t_1,t_2,t_3\dots)$ are the standard Schur polynomials defined by the
formula $\exp({\sum_{n=1}^\infty z^n t_n})=\sum_{m=0}^\infty z^m
P_m(t_1,t_2,t_3,\dots)$, and canonical basis $h^*_2,h^*_3,h^*_4,\dots$ given
by 
\begin{equation}  \label{bigcell-h*}
h^*_n=p^*_n-H^1_{n-1}, \qquad n=2,3,4,\dots,
\end{equation}
i.e. $h^*_n=p^*_n-u_{n-1}$, $n=2,3,4,\dots$ are the most convenient for this
purpose. In these coordinates one has the identity 
\begin{equation}
\{h^*_n,h^*_m\}=J^*_{n\ m-1}-J^*_{m\ n-1}, \qquad n,m=2,3,4,\dots
\end{equation}
where $J^*_{n m}$ denotes the Poisson tensor in these coordinates. So the
conditions $\{h_n,h_m\}\subset I(\Gamma_\infty)$ is satisfied if and only if
the conditions (\ref{bigcellJcond}) are valid. $\square$

On $\Gamma_\infty$ one has $p_n^*=u_{n-1}$, $n=2,3,4,\dots$ and, hence, 
\begin{equation}  \label{bigcellJ*=ab}
J^*_{ik}|_{\Gamma_\infty}=\alpha_{ik}(u)+\beta_{ik}(u)p^*_1, \qquad
i,k=1,2,3\dots
\end{equation}
where $\alpha_{ik}$ and $\beta_{ik}$ are functions of $u_k$ only. Since $%
p^*_1 \notin I(\Gamma_{\infty})$ the conditions (\ref{bigcellJcond}) are
equivalent to 
\begin{equation}  \label{bigcella=b=0}
\alpha_{i\ k-1}=\alpha_{k\ i-1}, \quad \beta_{i\ k-1}=\beta_{k\ i-1}, \qquad
i,k=1,2,3,\dots \ .
\end{equation}
The property (\ref{bigcellJ*=ab}) indicates that Poisson tensors $J^*_{ik}$
linear in the variables $p^*_k$ could be of particular relevance. Thus let
us consider the following class of tensors $J^*_{ik}$ 
\begin{equation}  \label{bigcellJ*}
J^*_{lk}=-\sum_m\frac{1}{m} J_{mk}(u)p^*_{l-m}
\end{equation}
where $J_{mk}(u)$ depend only on the variables $u_1,u_2,u_3,\dots$ . The
conditions (\ref{bigcellJcond}) or (\ref{bigcellJ*=ab}) are equivalent to
the following 
\begin{equation}  \label{bigcellJcond2}
\begin{split}
& \frac{1}{m}J_{mn}-\frac{1}{n}J_{nm}=0, \\
& \frac{1}{m}J_{m\ n-1}-\frac{1}{n}J_{n\ m-1} + \sum_{k=1}^{m-2}\frac{1}{k}%
u_{m-k-1}J_{k\ n-1} - \sum_{k=1}^{n-2}\frac{1}{k}u_{n-k-1}J_{k\ m-1}=0,
\quad n,m=1,2,3,\dots\ .
\end{split}%
\end{equation}
Using the well-known property of Schur's polynomials, i.e. $%
\partial_{p_k}P_n(p)=P_{n-k}(p)$ which implies that $\partial_{p_i} h =\-%
\frac{1}{i}\sum_{l=1}^{i-1}p^*_{l-i}\partial_{p_l^*}h$, one easily concludes
that the Poisson structure (\ref{py=J}) with $J^*_{lk}$ of the form (\ref%
{bigcellJ*}) in the coordinates $p_1,p_2,\dots;u_1,u_2,\dots$ has the form 
\begin{equation}  \label{pu=Ju}
\{p_i,p_k\}=0, \quad \{u_i,u_k\}=0, \quad \{u_i,p_k\}=J_{ki}(u), \qquad
i,k=1,2,3,\dots\ .
\end{equation}

\begin{obs}
The system (\ref{bigcellJcond2}) is equivalent to the system (\ref%
{bigcelllin}) modulo the associativity conditions (\ref{bigcellsercoeff})
and $J_{nm}=\Delta _{nm}$. So there is a strong interrelation between
cohomological and Poisson structures associated with the subbundle $%
TB_{W_{\varnothing }}$ for the big cell.
\end{obs}

This fact has been checked by computer calculations up to $n,m=11$. We do
not have formal proof of this statement.

Note that due to the properties of the Schur polynomials the Poisson tensor (%
\ref{pu=Ju}) is of the form 
\begin{equation}  \label{bigcellJJ*}
J^*_{ik}=\sum_{m} J_{mk}(u)\frac{\partial p_i^*}{\partial p_m}.
\end{equation}
A subclass of the Poisson tensors (\ref{bigcellJJ*}) for which $J_{mk}(u)=%
\frac{\partial u_k}{\partial x_m}$, i.e. 
\begin{equation}  \label{bigcellJJ*-sub}
J^*_{ik}=\sum_{m} \frac{\partial p_i^*}{\partial p_m}\frac{\partial u_k}{%
\partial x_m} , \qquad i,k=1,2,3,\dots
\end{equation}
where $x_1,x_2,x_3,\dots$ are new coordinates on $\mathcal{M}$, is of
particular interest. First in the coordinate $x_i$, $p_i$ the Poisson
structures (\ref{py=J}), {\ref{pu=Ju}} take the form 
\begin{equation}  \label{py=Darb}
\{p_i,p_k\}=0, \quad \{x_i,x_k\}=0, \quad \{x_i,p_k\}=\delta_{ki}, \qquad
i,k=1,2,3,\dots\ .
\end{equation}
i.e., the coordinates $p_i,x_i$, $i=1,2,3,\dots$ are the Darboux coordinates
in $\mathcal{M}$. Second, the Jacobi conditions (\ref{bigcell-Jacobi}) are
identically satisfied for the Ansatz $J_{ik}(u)=\frac{\partial u_k}{\partial
x_i}$ while the algebraic constraints (\ref{bigcellJcond2}) become the
system of quasilinear equations 
\begin{equation}  \label{bigcelluxcond}
\begin{split}
& \frac{1}{m}\frac{\partial u_n}{\partial x_m}-\frac{1}{n} \frac{\partial u_m%
}{\partial x_n} =0, \\
& \frac{1}{m}\frac{\partial u_{n-1}}{\partial x_m}-\frac{1}{n}\frac{\partial
u_{n-1}}{\partial x_m} + \sum_{k=1}^{m-2}\frac{1}{k}u_{m-k-1} \frac{\partial
u_{n-1}}{\partial x_k} - \sum_{k=1}^{n-2}\frac{1}{k}u_{n-k-1} \frac{\partial
u_{m-1}}{\partial x_k}=0, \quad n,m=1,2,3,\dots\ .
\end{split}%
\end{equation}
This system of equations coincides with that derived in \cite{KM2} in a
different manner. It was shown in \cite{KM2} that the system (\ref%
{bigcelluxcond}) is equivalent to the dKP hierarchy. This fact provide us
with an alternative proof of the Proposition \ref{prop-bigcellansatz}.

Thus we have

\begin{obs}
In the Darboux coordinates the system of equations (\ref{pu=Ju}), (\ref%
{bigcellJcond2}) characterizing the Poisson structure for the family of
ideals $I(\Gamma_{\infty})$ is equivalent to the dKP hierarchy with $%
x_1,x_2,x_3,\dots$ and $u_1,u_2,u_3,\dots$ playing the role of independent
and dependent variables, respectively.

The sets of variables $(p^*_k,u_k)$ and $(p^*_k,x_k)$ play the dual roles in
the description of the families of ideals $I(\Gamma_{\infty})$. The former
are canonical from the algebraic viewpoint while the latter are canonical
within the interpretation of the family of ideals $I(\Gamma_{\infty})$ as
Poisson ideal. In virtue of the formulas (\ref{bigcellJcond2}),(\ref%
{bigcellJJ*-sub}) the connection between these two sets of variables is
provided by solutions of the dKP hierarchy.
\end{obs}

This observation points out the deep interrelation between the theory of
Poisson ideals for the families of algebraic curves in Sato Grassmannian and
theory of integrable hierarchies and the role of Darboux coordinates in such
interconnection. The variables $x_k,\ k=1,2,3,\dots$ are deformation
parameters within such an approach. They can be viewed as the local
coordinates in the infinite-dimensional base space for coisotropic
deformations of the associative algebra (\ref{alg}).

The Darboux coordinates has been used in \cite{KM} within the study of
coisotropic deformations of the relations (\ref{alg},\ref{ass}) viewed as
equations defining structure constants of associative algebras. It was shown
in \cite{KM} that for infinite-dimensional polynomial algebra in the Fa\`{a}
di Bruno basis for which structure constants $C^l_{jk}$ are given by (\ref%
{bigcellstructcoeff}) the coisotropy condition (\ref{IGIG=IG}) is equivalent
to the associativity conditions (\ref{bigcellsercoeff}) plus the exactness
conditions 
\begin{equation}  \label{bigcell-exacta}
\frac{\partial H^i_n}{\partial x_l}=\frac{\partial H^l_n}{\partial x_i},
\qquad i,l,n=1,2,3,\dots \ .
\end{equation}
These conditions together with the algebraic relations $nH^i_n=iH^n_i$ imply
the existence of a function $F$ such that \cite{KM} 
\begin{equation}
H^i_m=-\frac{1}{m}\frac{\partial^2 F}{\partial x_i\partial x_m}.
\end{equation}
With such a form of $H^i_k$ the associativity conditions (\ref%
{bigcellsercoeff}) are equivalent to the celebrated Hirota-Miwa bilinear
equations (\ref{Hirota}).

This result indicates one more time the importance of the Darboux
coordinates in the whole our approach. The detailed analysis of the Poisson
structures for ideals of the families of algebraic curves in Birkhoff strata
and their connection with the hierarchy of integrable equations will be
given in the forthcoming paper.

Here we will present an illustrative example. In the subbundle $TB_{W_{1}}$
one has the ideal 
\begin{equation}
I(\Gamma _{\infty }^{1})=\langle \mathcal{F}%
_{23}^{1},h_{4}^{(1)},h_{5}^{(1)},\dots \rangle  \label{1strat-I}
\end{equation}%
where

\bigskip 
\begin{equation}
\begin{split}
\mathcal{F}_{23}^{1}=& {p_{3}}^{2}-{p_{2}}^{3}-\mu _{4}p_{2}p_{3}-\mu _{3}{%
p_{2}}^{2}-\mu _{2}p_{3}-\mu _{2}p_{2}-\mu _{0}, \\
h_{4}^{(1)}=& p_{4}-{p_{2}}^{2}-v_{2}p_{3}-v_{1}p_{2}-v_{0}
\end{split}%
\end{equation}%
and so on ( see \cite{KO1}).

\bigskip The requirement that the family of ideals (\ref{1strat-I}) is a
Poisson ideal gives rise to an infinite hierarchy of systems of PDEs. The
simplest of them which is equivalent to the condition $\{\mathcal{F}%
_{23}^{1},h_{4}^{(1)}\}\Big{|}_{\Gamma _{\infty }^{1}}=0$ with the canonical
Poisson bracket is given by (see also \cite{KO}) {\small 
\begin{equation}
\begin{split}
{\frac{\partial \mu _{{4}}}{\partial x_{4}}}=& -\frac{2}{3}\frac{\partial }{%
\partial x_{2}}(\mu _{2}\mu _{3})-\frac{5}{9}{\mu _{4}}^{2}\frac{\partial
\mu _{4}}{\partial x_{2}}+\frac{4}{9}\mu _{4}\frac{\partial \mu _{4}}{%
\partial x_{3}}+2\frac{\partial \mu _{2}}{\partial x_{2}}+\frac{4}{3}\frac{%
\partial \mu _{3}}{\partial x_{3}}+\frac{4}{9}\frac{\partial \mu _{4}}{%
\partial x_{2}}{\frac{\partial }{\partial x_{2}}}^{-1}\frac{\partial \mu _{4}%
}{\partial x_{3}}+\frac{8}{9}{\frac{\partial }{\partial x_{2}}}^{-1}{\frac{%
\partial ^{2}\mu _{4}}{\partial {x_{3}}^{2}}}, \\
{\frac{\partial \mu _{{3}}}{\partial x_{4}}}=& -\frac{2}{3}\mu _{{4}}\mu _{{3%
}}{\frac{\partial \mu _{{4}}}{\partial x_{2}}}+v_{{1}}{\frac{\partial \mu _{{%
3}}}{\partial x_{2}}}+2\,{\frac{\partial \mu _{{1}}}{\partial x_{2}}}-3\,{%
\frac{\partial v_{{0}}}{\partial x_{2}}}-2\,\mu _{{3}}{\frac{\partial v_{{1}}%
}{\partial x_{2}}}+\frac{2}{3}\mu _{4}{\frac{\partial \mu _{{3}}}{\partial
x_{3}}}-\mu _{{4}}{\frac{\partial v_{{1}}}{\partial x_{3}}}+\frac{4}{3}\,\mu
_{{3}}{\frac{\partial \mu _{4}}{\partial x_{3}}}, \\
{\frac{\partial \mu _{{2}}}{\partial x_{4}}}=& -\frac{2}{3}\mu _{{4}}\mu _{{2%
}}{\frac{\partial \mu _{{4}}}{\partial x_{2}}}+2\,{\frac{\partial v_{{0}}}{%
\partial x_{3}}}+v_{{1}}{\frac{\partial \mu _{{2}}}{\partial x_{2}}}-\mu _{{4%
}}{\frac{\partial v_{{0}}}{\partial x_{2}}}-\frac{2}{3}\mu _{{1}}{\frac{%
\partial \mu _{{4}}}{\partial x_{2}}}+\frac{2}{3}\mu _{{4}}{\frac{\partial
\mu _{{2}}}{\partial x_{3}}}+\frac{2}{3}\mu _{{2}}{\frac{\partial \mu _{{4}}%
}{\partial x_{3}}}, \\
{\frac{\partial \mu _{{1}}}{\partial x_{4}}}=& -\frac{2}{3}\mu _{{4}}\mu _{{1%
}}{\frac{\partial \mu _{{4}}}{\partial x_{2}}}+2\,{\frac{\partial \mu _{{0}}%
}{\partial x_{2}}}+v_{{1}}{\frac{\partial \mu _{{1}}}{\partial x_{2}}}%
-2\,\mu _{{3}}{\frac{\partial v_{{0}}}{\partial x_{2}}}-\mu _{{1}}{\frac{%
\partial v_{{1}}}{\partial x_{2}}}+\frac{2}{3}\mu _{{4}}{\frac{\partial \mu
_{{1}}}{\partial x_{3}}}-\mu _{{4}}{\frac{\partial v_{{0}}}{\partial x_{3}}}%
-\mu _{{2}}{\frac{\partial v_{{1}}}{\partial x_{3}}}+\frac{4}{3}\,\mu _{{1}}{%
\frac{\partial \mu _{{4}}}{\partial x_{3}}}, \\
{\frac{\partial \mu _{{0}}}{\partial x_{4}}}=& v_{{1}}{\frac{\partial \mu _{{%
0}}}{\partial x_{2}}}-\mu _{{1}}{\frac{\partial v_{{0}}}{\partial x_{2}}}+%
\frac{2}{3}\mu _{{4}}{\frac{\partial \mu _{{0}}}{\partial x_{3}}}-\mu _{{2}}{%
\frac{\partial v_{{0}}}{\partial x_{3}}}-\frac{2}{3}\mu _{{4}}\mu _{{0}}{%
\frac{\partial \mu _{{4}}}{\partial x_{2}}}+\frac{4}{3}\,\mu _{{0}}{\frac{%
\partial \mu _{{4}}}{\partial x_{3}}}
\end{split}
\label{KP-1}
\end{equation}%
} where $v_{1}=\frac{2}{3}\mu _{3}-\frac{2}{9}{\mu _{4}}^{2}+\frac{4}{3}{%
\frac{\partial }{\partial x_{2}}}^{-1}\frac{\partial \mu _{4}}{\partial x_{3}%
}$ and $v_{0}$ is associated with a gauge freedom of the system.

For the stratum $\Sigma_{1,2}$ the coisotropy condition (\ref{IGIG=IG}) is
given by pretty large system of equations. For example the condition 
\begin{equation}  \label{C8C9=0}
\{\mathcal{C}_8,\mathcal{C}_9\}\Big{|}_{\Gamma_{\infty}^2}=0
\end{equation}
with the Poisson bracket in the Darboux coordinates $(p_3,p_4,p_5,%
\dots,x_3,x_4,x_5,\dots)$ and cyclic variable $x_3$, is equivalent to the
system 
\begin{equation}
\partial_{x_4}H^5_i=\partial_{x_5}H^4_i, \qquad i=1,2,4,5
\end{equation}
where 
\begin{equation}
\begin{split}
H^4_4=&3\,U_{x_4}U_{x_5}V_{x_4} -2\,U_{x_4}{V_{x_4}}^{2}-H^4_{{2}}U_{x_4}
-2\,{U_{x_4}}^{2}V_{x_5}-U_{x_4}{U_{x_5}}^{2} +{U_{x_4}}^{4}-H^4_{{1}%
}U_{x_5} +{V_{x_5}}^{2} +V_{x_4}H^4_{{1}}, \\
H^4_5=&2\,H^4_{{2}}V_{x_4}-6\,U_{x_5}{V_{x_4}}^{2} +5\,V_{x_4}{U_{x_5}}^{2}-%
\frac{5}{3}\,V_{x_4}{U_{x_4}}^{3} -2\,U_{x_5}H^4_{{2}}+\frac{4}{3}\,U_{x_5}{%
U_{x_4}}^{3} -H^4_{{1}}V_{x_5} \\
& +3\,V_{x_4}U_{x_4}V_{x_5} -2\,U_{x_5}U_{x_4}V_{x_5}+\frac{7}{3}\,{V_{x_4}}%
^{3}-\frac{4}{3}\,{U_{x_5}}^{3}, \\
H^5_1=&{V_{x_4}}^{2}+2\,H^4_{{2}}+U_{x_4}V_{x_5} -2\,U_{x_5}V_{x_4}+{U_{x_5}}%
^{2}-{U_{x_4}}^{3}, \\
H^5_2=&2\,H^4_{{1}}U_{x_4}-V_{x_5}V_{x_4} +V_{x_5}U_{x_5}-V_{x_4}{U_{x_4}}%
^{2}, \\
H^5_4=&2\,H^4_{{2}}V_{x_4}-7\,U_{x_5}{V_{x_4}}^{2} +6\,V_{x_4}{U_{x_5}}^{2}-%
\frac{4}{3}\,V_{x_4}{U_{x_4}}^{3} -2\,U_{x_5}H^4_{{2}}+\frac{5}{3}\,U_{x_5}{%
U_{x_4}}^{3} -H^4_{{1}}V_{x_5} \\
& +4\,V_{x_4}U_{x_4}V_{x_5} -3\,U_{x_5}U_{x_4}V_{x_5}+\frac{8}{3}\,{V_{x_4}}%
^{3} -\frac{5}{3}\,{U_{x_5}}^{3}-H^4_{{1}}{U_{x_4}}^{2}, \\
H^5_5=&5\,U_{x_5}V_{x_4}{U_{x_4}}^{2}-2\,{V_{x_4}}^{2}{U_{x_4}}^{2} -3\,H^4_{%
{2}}{U_{x_4}}^{2}-4\,{U_{x_4}}^{3}V_{x_5} -2\,{U_{x_4}}^{2}{U_{x_5}}^{2}+2\,{%
U_{x_4}}^{5} \\
& +2\,U_{x_4}{V_{x_5}}^{2}-2\,V_{x_4}H^4_{{1}}U_{x_4} +{H^4_{{1}}}%
^{2}+2\,V_{x_5}{V_{x_4}}^{2} -3\,V_{x_4}V_{x_5}U_{x_5}+V_{x_5}{U_{x_5}}^{2}
+H^4_{{2}}V_{x_5}
\end{split}%
\end{equation}
and $\partial_{x_4}H^5_{-2}=\partial_{x_5}H^4_{-2}:=\partial_{x_4}%
\partial_{x_5}U$ and $\partial_{x_4}H^5_{-1}=\partial_{x_5}H^4_{-1}:=%
\partial_{x_4}\partial_{x_5}V$.

Finally, the requirement that the family of ideals $I^2(\Gamma^2_{\infty})$
for hyperelliptic curves is the Poisson ideal with respect to the canonical
Poisson bracket gives rise to the infinite hierarchy of hydrodynamical type
systems which is equivalent to that found in the paper \cite{KK}.


\subsubsection*{Acknowledgments}

The authors thank Marco Pedroni and Andrea Previtali for many fruitful
discussions. The author also thanks the referees for useful suggestions.
This work has been partially supported by PRIN grant no
28002K9KXZ and by FAR 2009 (\emph{Sistemi dinamici Integrabili e Interazioni
fra campi e particelle}) of the University of Milano Bicocca.

\end{document}